\documentclass[traditabstract,usenatbib]{aa}

\usepackage[varg]{txfonts}

\usepackage{mathtools}
\usepackage{aas_macros}
\usepackage{graphicx}
\usepackage{amstext}
\usepackage{enumitem}
\usepackage{amssymb}
\usepackage{booktabs}
\usepackage{grffile}
\usepackage{multicol}

\usepackage{xcolor}

\usepackage{subcaption}

\begin{document}

\newcommand{\mathleft}{\@fleqntrue\@mathmargin0pt}
\newcommand{\mathcenter}{\@fleqnfalse}
\newcommand{\unit}[1]{\ensuremath{\, \mathrm{#1}}}
\providecommand{\e}[1]{\ensuremath{\times 10^{#1}}}
\providecommand*{\printsecond}[2]{#2}
\newcommand{\edited}[1]{{#1}}
\providecommand{\ang}[1]{\ensuremath{{#1}^{\circ}}}

\newcommand{\figwidtha}{0.0375}
\newcommand{\figwidthb}{0.057}
\newcommand{\figwidthc}{0.066}

\newcommand{\figgapa}{0.541cm}
\newcommand{\figgapb}{0.185cm}
\newcommand{\figgapc}{0.02cm}

\newcommand{\figwidthdoublea}{0.075}
\newcommand{\figwidthdoubleb}{0.114}
\newcommand{\figwidthdoublec}{0.132}

\newcommand{\figgapdoublea}{1.042cm}
\newcommand{\figgapdoubleb}{0.330cm}
\newcommand{\figgapdoublec}{0.00cm}

\newcommand{\figwidthsurfdoublea}{0.118}
\newcommand{\figwidthsurfdoubleb}{0.132}
\newcommand{\figwidthsurfdoublec}{0.132}

\newcommand{\figgapsurfdoublea}{0.218cm}
\newcommand{\figgapsurfdoubleb}{0.000cm}
\newcommand{\figgapsurfdoublec}{0.000cm}

\newcommand{\LHlabelwidth}{0.1}

\newcommand{\figwidthdoublebx}{0.0674}
\newcommand{\figwidthdoublecx}{0.0792}

\title{Suppression of lithium depletion in young low-mass stars from fast rotation}

\titlerunning{Suppression of lithium depletion in young low-mass stars from fast rotation}

\author{T. Constantino \inst{\ref{inst1}}
        \and
        I. Baraffe \inst{\ref{inst1},\ref{inst2}}
        \and
        T. Goffrey \inst{\ref{inst3}}
        \and
        J. Pratt \inst{\ref{inst4}}
        \and
        T. Guillet \inst{\ref{inst1}}
        \and
        D. G. Vlaykov \inst{\ref{inst1}}
        \and
        L. Amard \inst{\ref{inst1}}
        }

  \institute{Physics and Astronomy, University of Exeter, Exeter, EX4 4QL, United Kingdom, \email{T.Constantino@exeter.ac.uk} \label{inst1} 
      \and
      Ecole Normale Sup\'erieure de Lyon, CRAL, UMR CNRS 5574, 69364 Lyon Cedex 07, France \label{inst2} 
       \and
      Centre for Fusion, Space and Astrophysics, Department of Physics, University of Warwick, Coventry, CV4 7AL, UK \label{inst3} 
      \and
      Department of Physics and Astronomy, Georgia State University, Atlanta GA 30303, USA \label{inst4} 
      }

\authorrunning{Constantino et al.}

\abstract{We compute rotating 1D stellar evolution models that include a modified temperature gradient in convection zones and criterion for convective instability inspired by rotating 3D hydrodynamical simulations performed with the {\sc music} code.  
In those 3D simulations we found that convective properties strongly depend on the Solberg-H{\o}iland criterion for stability.  We therefore incorporated this into 1D stellar evolution models by replacing the usual Schwarzschild criterion for stability and also modifying the temperature gradient in convection zones.
We computed a grid of 1D models between 0.55 and 1.2 stellar masses from the pre-main sequence to the end of main sequence in order to study the problem of lithium depletion in low-mass main sequence stars.  This is an ideal test case because many of those stars are born as fast rotators and the rate of lithium depletion is very sensitive to the changes in the stellar structure.  Additionally, observations show a correlation between slow rotation and lithium depletion, contrary to expectations from standard models of rotationally driven mixing.  
By suppressing convection, and therefore decreasing the temperature at the base of the convective envelope, lithium burning is strongly quenched in our rapidly rotating models to an extent sufficient to account for the lithium spread observed in young open clusters.
}


\maketitle

\section{Introduction}
\label{sec:lithium_and_rotation}

Lithium is a valuable tracer for constraining stellar evolution models.  $^7$Li is destroyed at temperatures above about 2.5 million K and it therefore provides a strong indication of the depth of the circulation of the surface material.  It has been used to probe mixing in main sequence stars and deep mixing in red giants \citep[e.g.][]{2007A&A...467L..15C,2010A&A...522A..10C,2015MNRAS.450.2423A}.  Additionally, the lithium depletion boundary (LDB) technique is used to determine the ages of open clusters with ages between about 20 and 200 Myr \citep{1996ApJ...458..600B,2004ApJ...604..272B}.

Over the last few decades, observations have highlighted the relationship between rotation and the surface lithium abundance in low-mass main sequence stars.  
\citet{1987ApJ...319L..19B} found that rapidly rotating Pleiades early-K dwarfs were more lithium rich than their slowly rotating counterparts. In a larger sample of F, G, and K dwarfs in the younger $\alpha$~Persei cluster, \citet{1988ApJ...333..267B} then detected the same trend.  
\citet{1988A&A...201..267R} identified a slight rotation-lithium trend in F and G stars in the much older Hyades cluster.  
\citet{1993AJ....106.1059S} then again found the correlation between fast rotation and lithium abundance in a larger sample of Pleiades G and K dwarfs and suggested that fast rotation is a direct cause of reduced lithium depletion.  
\citet{1997AJ....114..352J} next saw the same trend in the 250\,Myr old cluster M34 and \citet{1998MNRAS.300..550J} similarly found the trend for the roughly 150\,Myr old cluster NGC 2516.  
\citet{2000AJ....119..859K} did not find a one-to-one relation between lithium and photometric rotation periods in the Pleiades and therefore argued it was unlikely that rotation was responsible for the lithium spread, instead favouring problems with radiative transfer calculations, small metallicity variations, or  magnetic fields.  
\citet{2001A&A...377...84T} detected a lithium-photometric rotation period trend in solar-type stars in the Pleiades but not in the young cluster IC~2602, IC~2391, and $\alpha$~Persei.    
\citet{2007NewA...12..265X} and later \citet{2010ApJ...723.1542X} reported that among G and K Weak-line T Tauri pre-main-sequence (PMS) stars between 0.9\,M$_\odot$ and 1.4\,M$_\odot$ in the in Taurus-Auriga Nebula, higher surface lithium correlates with slower rotation, but this has been explained as an age effect.  The rotation-lithium pattern in the Pleiades is closely mirrored in the similarly aged cluster M35 \citep{2021MNRAS.500.1158J}.  \citet{2016A&A...590A..78B} found that there is already a link between rotation and lithium in cool dwarfs in the 5\,Myr star forming region NGC 2264. \citet{2016A&A...596A..29M} reported a strong correlation between rotation and lithium equivalent width for members of the $24 \pm 3$\,Myr $\beta$\,Pictoris association \citep{2015MNRAS.454..593B}.  The rotation-lithium trend is a widespread phenomenon in young clusters and is apparent from very young ages.  In older clusters, the trend may also exist but be obscured by the convergence of rotation rates.

That rotation appears to suppress mixing is at odds with expectations from standard models with rotational mixing:  Faster rotating models should experience more mixing and therefore more lithium depletion.  This can be seen, for example, in the evolution models from \citet{2005Sci...309.2189C} that match the range of lithium observed in clusters at various ages but where the lithium-rotation correlation is the opposite of that observed, as well as the models from \citealt{2016A&A...587A.105A} that included a number of different treatments for determining turbulent diffusion coefficients from \citet{1992A&A...265..115Z}, \citet{1997A&A...321..134M}, \citet{1997A&A...317..749T}, and \citet{2004A&A...425..243M}.  The hydrostatic effects from including the centrifugal force from rotation do not explain the observed trend, and the discrepancy is worsened when rotational mixing is included \citep{1999A&A...341..174M}.

Several mechanisms have been proposed to explain the counter-intuitive empirical correlation between lithium and rotation.  Without providing a physical justification, \citet{1994ASPC...64..211S} conjectured that changes in the temperature-density profile in the lower regions of the envelope may be responsible for the observed lithium-rotation correlation.  Many of the specific physical mechanisms later suggested are indeed those that cause radial inflation above standard non-rotating non-magnetic models because these changes will generally reduce the temperature at the base of the convection zone and therefore inhibit surface lithium depletion.

\citet{2014ApJ...790...72S} computed models with different mixing length theory (MLT) parameters to replicate the radius dispersion seen for Pleiades stars at a given age and mass.  The resulting change in the structure during the PMS produces a lithium spread comparable to the observations, and \citet{2015MNRAS.449.4131S} later found similar results with models that include rotationally induced mixing.  Possible physical justifications for the increase in radius include the presence of spots \citep[e.g.][]{1986A&A...166..167S,2014MNRAS.441.2111J} or a strong magnetic field inhibiting convection \citep[e.g.][]{2007A&A...472L..17C,2010ApJ...723.1599M,2013ApJ...779..183F,2015ApJ...807..174S,2015MNRAS.449.4131S}, or magnetic pressure \citep{2001ApJ...559..353M}.  These explanations would be supported by an empirical correlation between radius and lithium.  Although a correlation between rotation and radius inflation has been reported for both the Pleiades \citep{2017AJ....153..101S} and M35 \citep{2021MNRAS.500.1158J}, there is a lot of scatter and it is much weaker than that between rotation and the lithium abundance, and uncertainties on empirical radii are generally relatively large \citep[e.g.][]{2004AJ....127.1029R,2017A&A...597A..63L,2019MNRAS.483.1125J}.  \citet{2021MNRAS.500.1158J} also found a correlation between radius inflation and Li below about 4800\,K, providing support for the idea that these structural changes are an important mechanism.  Despite these clusters being quite young, however, it is the stellar radii at even younger ages, when lithium depletion is fastest, that are far more important for the scenario where there is a close connection between radius and lithium depletion.

\citet{1998A&A...331.1011V} suggested that the lithium-rotation pattern could be produced by a combination of convective overshooting balanced by rotationally induced magnetic fields that inhibit convection.  In the ideal gas case with no molecular weight gradient, the magnetic field alters the stellar structure via the criterion for convective stability, which becomes
\begin{equation}
\nabla_\text{rad}  < \nabla_\text{ad} + \frac{B_v^2}{B_v^2 + \gamma P},
\end{equation}
where $\nabla_\text{rad}$ is the radiative temperature gradient, $\nabla_\text{ad}$ is the adiabatic temperature gradient, $B_v^2$ is the square of the vertical magnetic field strength (in rationalized Gaussian units with $c=1$), $\gamma = c_p/c_v$, and $P$ is the gas pressure \citep{1966MNRAS.133...85G,1968MNRAS.141..165M}.  A steeper temperature gradient is therefore required before the convective instability sets in.  The role of magnetic fields in lithium depletion has, however, been questioned: \citet{2018A&A...613A..63B} note that at young ages the lithium spread grows as the rotational distribution of stars widens, while magnetic field strength declines, which indicates instead that rotation is the likely culprit.  They also point out that low-mass stars tend to be near the saturated activity level during the PMS, meaning there is no clear rotation-activity relationship, which would be needed to explain the later rotation-lithium correlation if magnetic field strength controls the lithium depletion rate.  Discussing their study of M35, \citet{2021MNRAS.500.1158J} note that stars currently with moderately slow 10~day periods would have had Rossby numbers low enough for the magnetic activity indicators to be saturated during the rapid lithium depletion phase \citep[e.g.][]{1984A&A...133..117V,2003A&A...397..147P,2009ApJ...692..538R,2009MNRAS.399..888M,2011ApJ...743...48W,2011MNRAS.411.2099J,2020AJ....159...52M}.  This suggests that even if magnetic effects tightly correlate with rotation, they alone cannot explain the observed lithium-rotation trend.  \citet{2021MNRAS.500.1158J} thus propose a scenario in which saturated magnetic activity diminishes the PMS Li depletion fairly uniformly and then the observed pattern is produced by rotation differentially inhibiting mixing beneath the convection zone.

Other proposed explanations relate more directly to the rotational properties.  In models of young stars with a slowly rotating surface due to a longer PMS disk-locking phase, mixing could be induced by more substantial differential rotation, leading to the observed correlation between slow surface rotation and Li depletion \citep{2008A&A...489L..53B,2012A&A...539A..70E}.  \citet{2014A&A...566A..72G} similarly suggested that the rotation-lithium links in the Pleiades and M34 may be explained by the slow rotators having experienced an episode of strong rotational braking that induces shear-driven mixing below the convection zone.  This echoes the earlier hypothesis from \citet{1997AJ....114..352J} that fast rotation preserves lithium during the PMS and that a high angular momentum loss rate speeds up its depletion.  \citet{2017ApJ...845L...6B} showed that a reasonable agreement between models and observations is possible with a simple rotationally dependent convective overshooting parameter.  The good news is that models with plausible physical inputs can reproduce the observed lithium pattern.  The challenge is determining which mechanisms are important.

\section{Stability criterion from 3D hydrodynamical simulations}

When rotation is present, the stability against convection is determined from the Solberg-H{\o}iland stability criteria \citep[e.g.][]{2000stro.book.....T}.  Firstly, 
\begin{equation}
\frac{1}{\varpi} \frac{\partial j^3}{ \partial \varpi}  - \boldsymbol{g} \cdot \left( \frac{1}{\Gamma_1 P} \nabla P - \frac{1}{ \rho }\nabla \rho \right)  > 0,
\label{Solberg1}
\end{equation}
where $\varpi = r \sin{\theta}$ is the distance from the rotation axis.  The second condition is
\begin{equation}
\frac{\partial P}{\partial z}  \left( \frac{\partial j^2 }{\partial \varpi} \frac{\partial s}{\partial z} - \frac{\partial j^2}{\partial z} \frac{\partial s}{\partial \varpi} \right) < 0,
\label{Solberg2}
\end{equation}
where $s$ is the specific entropy, $j$ is the specific angular momentum, and $ z = r \cos{\theta}$ is the distance from the equatorial plane.  The second criterion with solid body rotation, or more generally when $\frac{\partial j}{\partial z} = 0$, reduces to the usual Schwarzschild criterion except at the equator, so an instability will exist if $ \frac{\partial P}{\partial z} \frac{\partial s}{\partial z} > 0$ \citep[e.g.][]{1974A&A....33...99S}.  The situation is more complicated in real stars where there can be both a radial and latitude dependence of entropy and angular velocity, or where pressure and density gradients are misaligned.

We have computed a suite of 3D hydrodynamical simulations of rotating convection in solar-like envelopes using the Multi-Dimensional Stellar Implicit Code {\sc music} \citep[see e.g.][]{2016A&A...586A.153V,2017A&A...600A...7G}, which will be analysed in detail in a forthcoming paper.  Solar-like initial models were generated from a 1\,$M_\odot$ 1D stellar structure computed with the {\sc mesa} stellar evolution code that had metallicity $Z = 0.02$, luminosity $L = 1.04\,\text{L}_\odot$, radius $R = 1.018\,\text{R}_\odot$, and was approximately the solar age.  The 3D simulation domain includes the majority of the convective envelope $0.706 < r/R < 0.961$; three different colatitude extents (centred at the equator) $\pi/2$\,rad, $3\pi/4$\,rad and $7\pi/8$\,rad were tested; and the extent was $\pi/2$ in the azimuthal direction.  There are a total of 192 grid cells in the radial direction, 128 in the azimuthal direction and between 256 and 448 in the latitudinal direction, depending on the extent of the domain.  In the radial direction, cells are spaced so that the sound crossing time of each cell is uniform.  The Coriolis force is included and the runs were started with 0, 0.5, 1, 2, 3, and 5 times the solar rotation rate.

A reduction in convective velocity in the presence of rotation has been reported for 3D simulations before \citep[e.g.][]{2017ApJ...836..192B}.  Rather than depending solely on the usual measure of convective stability, the Brunt--V{\"a}is{\"a}l{\"a} frequency $N$, where
\begin{equation}
\label{eq_Brunt-Vaisala_frequency}
N^2 = \frac{g}{\rho}\left[ \left( \frac{\text{d}\rho}{\text{d}r} \right)_\text{ad} - \frac{\text{d}\rho}{\text{d}r}  \right],
\end{equation}
we found that the convective velocity additionally depends on the rotation.  Specifically, there is a clear correlation between radial velocity magnitude $|{v_r}|$ and one of the Solberg-H{\o}iland stability criteria \citep{1978trs..book.....T,1978ApJ...220..279E}:
\begin{equation}
\label{eq_adjusted_Brunt-Vaisala_frequency}
\frac{g}{\rho}\left[ \left( \frac{\text{d}\rho}{\text{d}r} \right)_\text{ad} - \frac{\text{d}\rho}{\text{d}r}  \right] + \frac{1}{r^3} \frac{\text{d}}{\text{d}r}(r^2 \omega)^2 \ge 0
\end{equation}
or, alternatively,
\begin{equation}
N^2 + \frac{1}{r^3} \frac{\text{d}}{\text{d}r}(r^2 \omega)^2 \ge 0,
\label{modified_stability_N2}
\end{equation}
which was suggested as early as \citet{1942ApJ....95..454R} and \citet{Walen_1946}.  When $N^2 + \frac{1}{r^3} \frac{\text{d}}{\text{d}r}(r^2 \omega)^2$ increases in our simulations, the temporally and spatially averaged radial velocity magnitude $<{|v_r|}>$ decreases, whereas there is no such correlation between $N^2$ and $<{|v_r|}>$.  While criterion~\eqref{eq_adjusted_Brunt-Vaisala_frequency} is true only in the equatorial plane, we did not notice a significant dependence of velocity on latitude.  In any case, applying a factor of $\sin{\theta}$ to account for only the radial component perpendicular to the rotation axis, and taking the average over the sphere, the second term on the left-hand side is only reduced by a factor of $\pi/4$.  \citet{2009pfer.book.....M} suggested this stability criterion for 1D models but with the very similar factor of $\sqrt{2/3}$.  This formulation ignores instabilities that may arise parallel to the rotation axis, depending on $ \frac{\partial j}{\partial \varpi} $ in inequality~\eqref{Solberg2}.  \citet{1951ApJ...114..272C} argued that when displacements in all directions are considered, an instability exists whenever the temperature gradient exceeds the adiabatic gradient, but also suggested that rotation may still inhibit the energy transport by convection.  This approach is also supported by rotating 3D simulations of Rayleigh-B\'{e}nard convection which show the efficiency of convection is also reduced in the direction of the rotation axis \citep{1996DyAtO..24..237J}.

It is not feasible to compute the 3D simulations for long enough for the thermal structure to adjust to the rotation.  Therefore, in order to gauge how well inequality~\eqref{modified_stability_N2} explains the rotation dependence of convection in our simulations, we first calculate the entropy gradient required to cancel out the contribution from rotation and ensure neutral stability at the equator:
\begin{equation}
\frac{\text{d}s}{\text{d}r} = \frac{c_P}{g}\delta N^2 = \frac{c_P}{g} \frac{1}{r^3} \frac{\text{d}}{\text{d}r}(r^2 \omega)^2,
\end{equation}
where we define $\delta N^2 = \frac{1}{r^3} \frac{\text{d}}{\text{d}r}(r^2 \omega)^2$, which is the second term in inequality~\eqref{modified_stability_N2}.
We can then integrate this to find
\begin{equation}
\Delta s_\text{rot} = -\int_{r_0}^{r_1} { \frac{c_P}{r^3 g} \frac{\text{d}}{\text{d}r}(r^2 \omega)^2   \text{d}r }
\label{eq_entropy_integral}
\end{equation}
and compare it to the entropy perturbation near the surface $\Delta s_\text{pert}$, which we define as a multiple of the standard deviation of the entropy probability density function at a depth near the surface at $r = r_1$ (and we chose a factor of 1.45 for a good fit).  The bounds of the integration, $r_0 = 5.25\times 10^{10}\,\text{cm}$ and $r_1 = 6.75\times 10^{10}\,\text{cm}$, contain 90 per cent of the convection zone for the non-rotating model ($ 5.16\times 10^{10}\,\text{cm} \lesssim r  < 6.8\times 10^{10}\, \text{cm}$) and exclude the very bottom of the convection zone and the near-surface boundary layer.   The lower part of the convection zone was excluded because the boundary location is simulation-dependent and the upper simulation boundary region was excluded so that the integration ends at a depth where plumes are properly formed, but the results hold independently of the exact choices.  The two quantities $\Delta s_\text{rot}$ and $\Delta s_\text{pert}$ should correspond to each another if the entropy perturbations are primarily generated by cooling at the surface and then plumes descend nearly adiabatically until they become more buoyant than their surroundings.  This assumption is justified by the dominant length scale of the velocity field being relatively large compared with the overall depth of the convection zone in our simulations, that is to say, coherent structures travel significant distances before they are dispersed.  The outermost layers of the star, near the photosphere, where we might expect to find smaller-scale structures, are also excluded from the simulation.  Radiation losses are not significant either: the radiative diffusion time across grid cells in the horizontal direction is greater than 100\,yr, which is much greater than the convective turnover (and simulation) time.

\begin{figure}
\includegraphics[width=\linewidth]{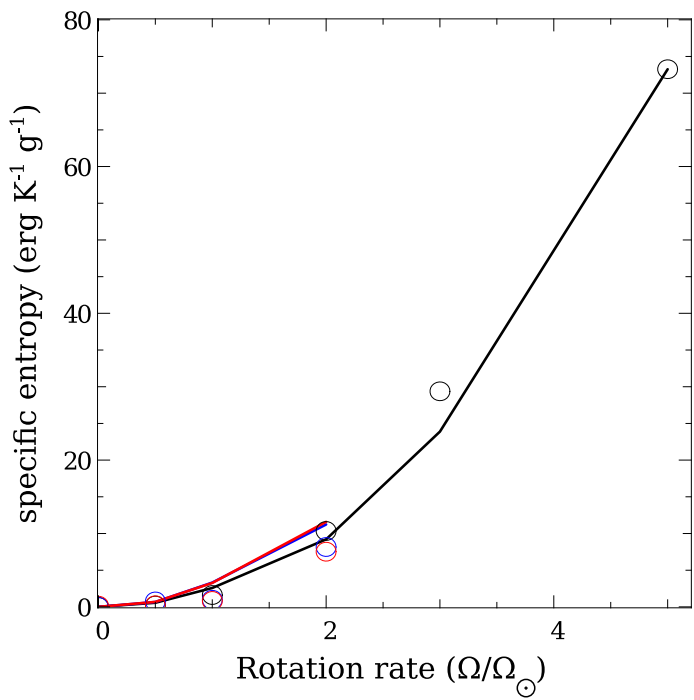}
  \caption{Comparison between the entropy integral $\Delta s _\text{rot}$ from Eq.~\eqref{eq_entropy_integral} denoted by lines and the near-surface entropy perturbation $\Delta s _\text{pert}$ (circles) from 3D rotating {\sc music} models.  $\Delta s_\text{pert}$ is defined as 1.45 times the standard deviation of the entropy probability density function at $r = r_1 = 6.75\times 10^{10}\,\text{cm}$, which is at a depth equivalent to three per cent of the convection zone.  The simulations have colatitude extents (centred at the equator) of  $\pi/2$\,rad (black), $3\pi/4$\,rad (blue), and $7\pi/8$\,rad (red), respectively.  In each case, $\Delta s_\text{pert}$ is shown after subtracting the entropy perturbation from a non-rotating model for the sake of clarity.}
  \label{figure_entropy_perturbation}
\end{figure}

Figure~\ref{figure_entropy_perturbation} demonstrates there is a tight correlation between the entropy integral $\Delta s _\text{rot}$ and $\Delta s_\text{pert}$ for all the models with a range of latitudinal extents and rotation rates.  This result is exactly as expected if inequality~\eqref{modified_stability_N2} correctly describes the stability.  For the sake of clarity we subtract $\Delta s_\text{pert}$ for a non-rotating model from each measurement in Fig.~\ref{figure_entropy_perturbation}.

The {\sc music} simulations were limited to a few convective turnover times.  The data for Fig.~\ref{figure_entropy_perturbation} was computed over 2 to 12 convective turnovers where the global turnover time is given by
\begin{equation}
\tau_g = \int_{r_0}^{r_1}{\frac{\text{d}r}{<|v_r|>}},
\end{equation}
and $<|v_r|>$ is the spatially and temporally averaged radial velocity magnitude and
$r_0 = 5.25 \times 10^{10}\,\text{cm}$ and $r_1 = 6.75 \times 10^{10}\,\text{cm}$, which spans almost the entire convection zone.  The relatively short duration of the 3D simulations means that in order to explore whether the above trends would affect predictions for the evolution of stars, 1D stellar evolution models must be used.  We selected the problem of lithium depletion in young low-mass stars because (i) there is a clear empirical rotation dependence, (ii) lithium is a sensitive diagnostic of the internal structure, specifically the depth of the convection zone, (iii) many of these stars are fast rotators, which will enhance any effect, and (iv) the envelopes of these stars are very similar to those in the 3D simulations.

\section{Stellar models}
\label{sec:1dmodels}

The 1D stellar evolution models for this paper were computed with {\sc monstar} \citep[see e.g.][]{2008A&A...490..769C,2014ApJ...784...56C}.  We constructed a grid of models with mass between 0.55 and 1.2\,M$_\odot$, which covers a broad range of stars in which the rotation-lithium connection is observed, from the PMS to the end of the main sequence.  In these models, the chemical mixing near the base of the convection zone, which is responsible for the depletion of lithium, has little effect on the structure, so the surface lithium abundance evolution for different overshooting schemes is calculated with a post-processing code.

All of the models initially have the solar metal abundance according to \citet{2009ARA&A..47..481A} with metallicity $Z = 0.013$, and we chose a moderate Galactic helium enrichment $Y = 0.278$.  The initial lithium abundance was $A(\text{Li}) = 12 + \log[N(\text{Li})/N(\text{H})] = 3.3$ to match the most lithium-rich stars in young open clusters.  The MLT mixing length is the solar-calibrated value of 1.60\,$H_\text{p}$ in each model.  No microscopic diffusion is applied.  Metallicity is an important factor for the evolution of lithium \citep{2002ApJ...566..419P}, which we do not investigate in this paper, so we do not attempt to precisely match the models to the open clusters we include.  Nuclear reaction rates are taken from the NACRE II compilation \citet{2013NuPhA.918...61X}.

\subsection{Rotation model}

In this study we adopt a very simple treatment of rotation.  Rotation is assumed to be solid-body and angular momentum is lost by magnetic braking according to Kawaler’s law (\citealt{1988ApJ...333..236K}; see details in \citealt{1997A&A...326.1023B,2012A&A...546A.113V}), consistent with the models from \citet{2017ApJ...845L...6B}.  The evolution of the angular momentum is thus
\begin{equation}
    \frac{\text{d}J}{\text{d}t} = 
\begin{cases}
    -K \Omega^3 \left( \frac{R}{\text{R}_\odot} \right) ^{1/2}  \left( \frac{M}{\text{M}_\odot} \right) ^{-1/2}     ,& \text{if } \Omega < \Omega_\text{sat} \\
    -K \Omega_\text{sat}^2 \Omega  \left( \frac{R}{\text{R}_\odot} \right) ^{1/2}  \left( \frac{M}{\text{M}_\odot} \right) ^{-1/2},              & \text{otherwise},
\end{cases}
\label{eq_angmom}
\end{equation}
where $\Omega_\text{sat} = 14 \Omega_\odot$ and $K = 2.7 \times 10^{47}  \ \text{g} \ \text{cm}^{-2} \ \text{s} $, based on the calibrations for the solar rotation from \citet{1997A&A...326.1023B} and \citet{2012A&A...546A.113V}.  This rotation law was applied from the PMS when the stellar radius was $11\,\text{R}_\odot$.  Instead of modelling a disk-locking phase, we set the initial rotation rates so that at 20\,Myr the fast rotators have $1.85 < \log{\Omega/\Omega_\odot} < 2.0$ and the slow rotators have $0.58 < \log{\Omega/\Omega_\odot} < 1.0$.  The fast rotators are comparable to the 90\textsuperscript{th} percentile of the rotational distribution observed in stellar clusters while the latter is between the 25\textsuperscript{th} and 50\textsuperscript{th} percentile \citep{2015A&A...577A..98G}.  This approach is analogous to that of \citet{2016ApJ...829...32S}, who initialized their rotation at 13\,Myr rather than including a disk-locking phase.  The only way in which our stellar structure models are affected by the rotation is by altering the adiabatic temperature gradient and the criterion for stability against convection.  We neglect the reduction of effective gravity from the centrifugal force as well as additional sources of mixing such as meridional circulation or turbulence induced by rotational shear.  The aim is primarily to qualitatively understand how our modifications specifically affect lithium depletion.  Integrating the stability criterion into models that better account for rotation is reserved for future work.  In subsequent models it would also be useful to compare the beryllium evolution because it is destroyed at higher temperatures and therefore probes mixing at a lower depth \citep[e.g.][]{1999ApJ...511..466B}.  For the sake of clarity, in this paper we present two sets of models, which are representative of fast and slow rotators, respectively.

\subsection{Implementing the stability criterion}

The velocity of flows in the radial direction of the rotating 3D hydrodynamical models does not show a strong dependence on latitude, so we use the condition in inequality~\eqref{modified_stability_N2}, which is strictly valid at the equator, without modification.  When solid-body rotation at rate $\Omega$ is assumed, inequality~\eqref{modified_stability_N2} can be used to derive a new stability condition:
\begin{equation}
\label{eq_stability_criterion}
 \nabla_\text{rad} \leq \nabla_\text{ad} - \frac{4 P \Omega^2}{\rho g^2 } \left( \frac{\partial \ln{T}}{\partial \ln{\rho}} \right)_{P},
\end{equation}
which is the Schwarzschild criterion with the addition of the so-called Rayleigh discriminant on the right-hand side.  This is equivalent to the formulation arrived at (although not actually used) by \citet{2008A&A...479L..37M} for 2D models of a 20\,M$_\odot$ star, except for the additional $\sin{\theta}$ factor.
The factor of $\Omega^2$ suggests that any effects will be much stronger when rotation is fast and it obviously reduces back to the ordinary Schwarzschild criterion when there is no rotation.  Where it is used elsewhere in the stellar evolution code, the adiabatic temperature gradient $\nabla_\text{ad}$ is similarly adjusted by addition of the Rayleigh discriminant.

\subsection{Overshooting schemes}
\label{sec_mixing}

Evidence from solar twins (main sequence stars with nearly the same mass and metallicity as the Sun) and open clusters is consistent with there being a gradual depletion of lithium during main sequence \citep[e.g.][]{2005A&A...442..615S,2008A&A...489..677P,2010Ap&SS.328..193M,2016A&A...587A.100C}.  Similarly, lithium measurements for stars with mass, radii, and rotation periods derived from photometry from the \textit{Kepler} satellite show that stars that are more slowly rotating, and therefore generally older, are more lithium depleted \citep{2017A&A...602A..63B}.  This implies that some kind of deep mixing is required throughout the main sequence evolution.

We computed models with a number of schemes for mixing in formally stable regions, and where possible calibrated free parameters so that a solar model with the solar rotation rate has the solar lithium abundance at the age of the Sun:
\begin{enumerate}
\item Exponential decay of the diffusion coefficient proposed by \citet{1997A&A...324L..81H} based on the 2D hydrodynamical simulations from \citet{1996A&A...313..497F}:
\begin{equation}
D_\text{os}(z) = D_0 e^{-2z/(f_\text{os}H_\text{p})},
\label{eq_herwig}
\end{equation}
where $D_0 = vl/3$ is the diffusion coefficient near the convective boundary from MLT, $z$ is the distance from the convective boundary, $H_\text{p}$ is the pressure scale height, and the calibrated value for the free parameter is $f_\text{os} = 0.035$.  This is hereafter referred to as the `Herwig' scheme.
\item A uniform diffusion coefficient in radiative regions in the outer half of the model (by mass): $D_\text{uni} = 4.4 \times 10^3$\,cm$^2$\,s$^{-1}$.  This is used without a specific physical motivation in mind.  It is merely representative of the kind of slow mixing that may occur and that appears to be needed during the main sequence.  We refer to this scheme as `deep diffusion'.  This is hereafter referred to as the `Pratt' scheme.
\item A combination of the previous two schemes where the diffusion coefficient applied is the larger from the two: $D = \max{(D_\text{os},D_\text{uni})}$.  In this case $f_\text{os} = 0.02$ was arbitrarily chosen for the exponential decay scheme and then a solar model Li calibration yielded $D_\text{uni} = 2.4 \times 10^3$\,cm$^2$\,s$^{-1}$.
\item The scheme proposed by \citet{2017A&A...604A.125P} based on 2D {\sc music} convection simulations \citep[see also][]{2016A&A...593A.121P} and later supported by 3D simulations \citep{2020A&A...638A..15P}:
\begin{equation}
D_\text{os}(z) = D_0 \bigg \{1- \exp \bigg [ -\exp\left(- { { (r_{\rm B} - r) \over R} - \mu \over \lambda} \right) \bigg ] \bigg \},  
\end{equation}
where $\mu = 0.005$ from \citet{2017ApJ...845L...6B} and $\lambda = 0.0026$ is from a solar calibration, $r$ is the radial coordinate, $R$ is the stellar radius, and $r_B$ is the position of the convective boundary.
\item Rotationally dependent overshooting used by \citet{2017ApJ...845L...6B}, which reduces above a threshold rotation rate $\Omega_\text{t} = 5\,\Omega_\odot$ and where the overshooting distance (using the same diffusion coefficient as \citealt{2017A&A...604A.125P}) is limited to 1\,$H_\text{P}$ if $\Omega < \Omega_\text{t}$ and 0.1\,$H_\text{p}$ otherwise.  This is the only scheme with parameters that directly depend on the rotation rate.
\end{enumerate}
Finally, we also included a sixth set of models without any non-convective mixing, which could not be calibrated to match the solar lithium abundance.

The slow, uniform diffusion scheme is included as a crude emulation of the slow but deep mixing that is generally needed to match the lithium depletion inferred from observations of solar-twins at different ages.  Proposed mechanisms include theoretically predicted mixing induced by gravity waves \citep[e.g.][]{1996A&A...305..513M,2000A&A...354..943M,2004A&A...418.1051T}, mixing from various rotationally induced instabilities \citep[e.g.][]{1999ApJ...525.1032B,2009A&A...495..271D}, and convective overshooting \citep[e.g.][]{2012MNRAS.427.1441Z,2015A&A...579A.122A,2017ApJ...845L...6B,2019ApJ...881..103Z}.  Best-fit models for asteroseismology also provide some support for slow mixing in regions formally stable to convection \citep[e.g.][]{2015A&A...580A..27M,2016ApJ...823..130M,2020ApJ...899...38W}.  Other authors have adopted diffusion coefficients of the same order of magnitude as ours for lithium fitting.  \citet{2010ApJ...719...28D} showed that a diffusion coefficient of $2 - 4 \times 10^3$\,$\text{cm}^2 \text{s}^{-1}$ is required to be consistent with observations of solar-like stars, which is much less than the required rate of angular momentum diffusion to match the spin-down of solar analogues, suggesting another angular momentum transport mechanism is required.  \citet{2018MNRAS.481.4389J} boosted the diffusion coefficient from \citet{2018MNRAS.477.3845C} for a helioseismic fit to the Sun by a factor of 100 to $1.5\times 10^{4}$\,$\text{cm}^2 \text{s}^{-1}$ to achieve the solar lithium abundance.  Calibrating our deep diffusion scheme gives a diffusion coefficient similar enough to these earlier mixing models for it to be a reasonable example of their kind for this study.

\section{Structural effects from rotation}

In order to examine how the rotation alters the stellar structure, we computed several 1\,M$_\odot$ models evolved to the age of the Pleiades (125~Myr) with a range of rotation rates.  These were not evolved using the rotation law used for the lithium evolution calculations that follow in Sect.~\ref{sec:lithium_evolution_effects}.  The rotation rates were instead manually adjusted to produce a spread.  The results are summarized in Table~\ref{table_structure}.  As the rotation rate increases, the required temperature gradient for convection to carry the luminosity increases.  Consequently, the  models become slightly more inflated and cooler.  This is analogous to the structural effect of additional opacity.  The effects on global properties are fairly small for modest rotation rates: The model rotating at one quarter the break-up value has a radius a little under one per cent above that of a non-rotating model and the effective temperature is 26\,K cooler.  The modest effect on the global properties is consistent with the 2D model of a fast-rotating 20\,M$_\odot$ OB star with two envelope convective shells computed by \citet{2008A&A...479L..37M}.  In that model, the Solberg-H{\o}iland criterion increased $\nabla$ by less than 0.01 and gave very small changes to the structure, despite the fast rotation rate: $\Omega / \Omega_\text{crit} = 0.94$.

The conditions in the convection zone, however, are much more sensitive to the rotation.  By changing the location of the convective boundary, the same rotation as above reduces the mass inside the convection zone by more than 10 per cent and consequently also reduces the temperature at the base of the convective envelope $T_\text{bce}$ from 2.0\,MK to 1.9\,MK.  These trends continue for faster rotation rates, and moreover, these properties become increasingly sensitive to the rotation rate: Just as with the changes to the temperature gradient, the reduction in $T_\text{bce}$ is proportional to $\Omega^2$.  Table~\ref{table_structure} also gives an indication of the importance of neglecting the reduction of effective gravity from the centrifugal acceleration on the structure.  The $T_\text{bce}$ cools substantially while $g_\text{eff}$ is only reduced by a few per cent.  Stellar models that do account for the reduction in $g_\text{eff}$ from \citet{1999A&A...341..174M}, and rotate at about $4\times 10^{-4}\,\text{rad}\text{s}^{-1}$ at the ZAMS show small increase in $T_\text{bce}$ over non-rotating models, $\Delta \log{T} \approx 0.02$, and only a negligible increase during the PMS when $T_\text{bce}$ is highest and lithium depletion is fastest.  This is also consistent with the rotating models from \citet{2016A&A...587A.105A} where the additional chemical transport from rotation is the dominant factor for the lithium evolution.  

\begin{figure}
\includegraphics[width=\linewidth]{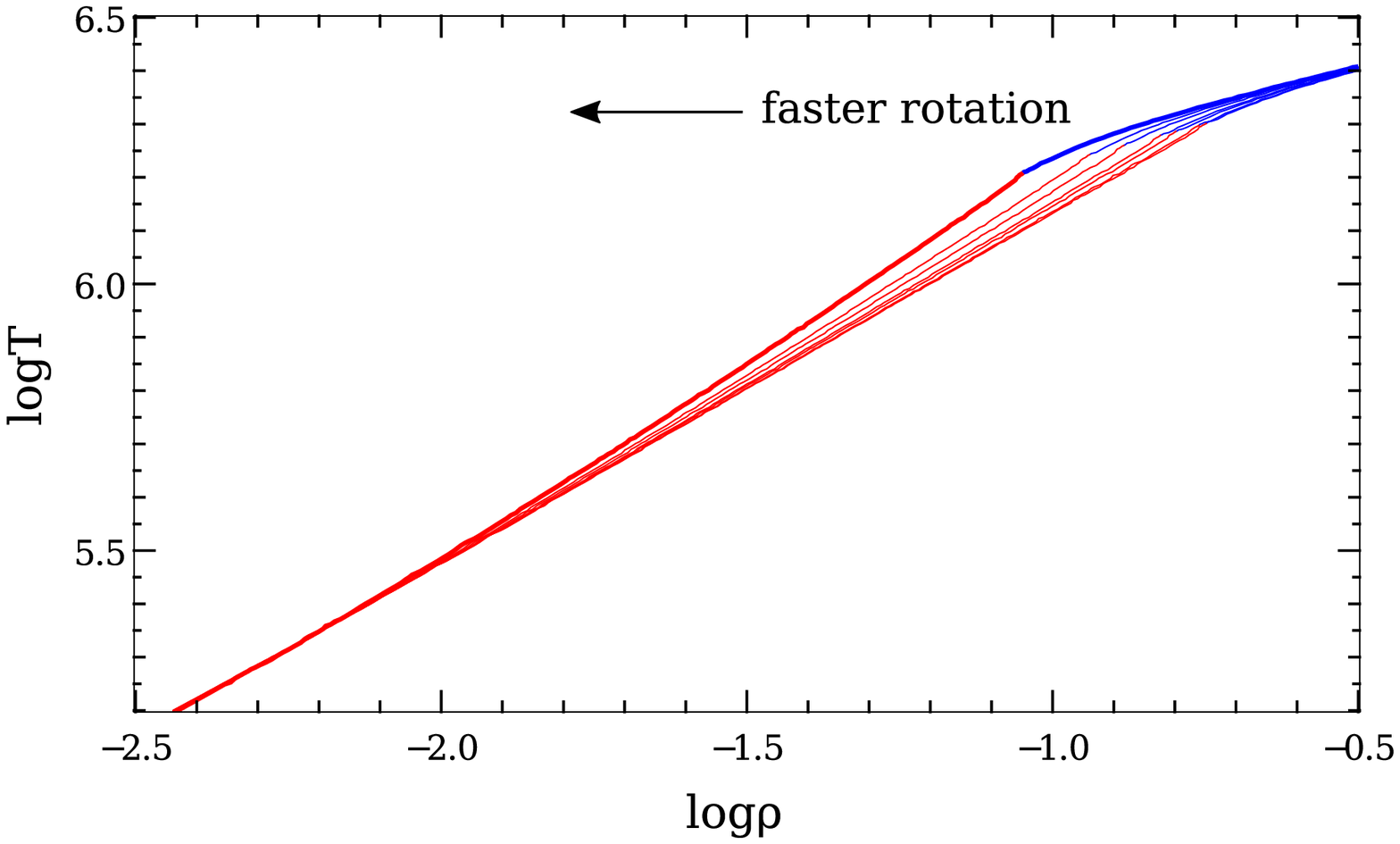}
\newline
\includegraphics[width=\linewidth]{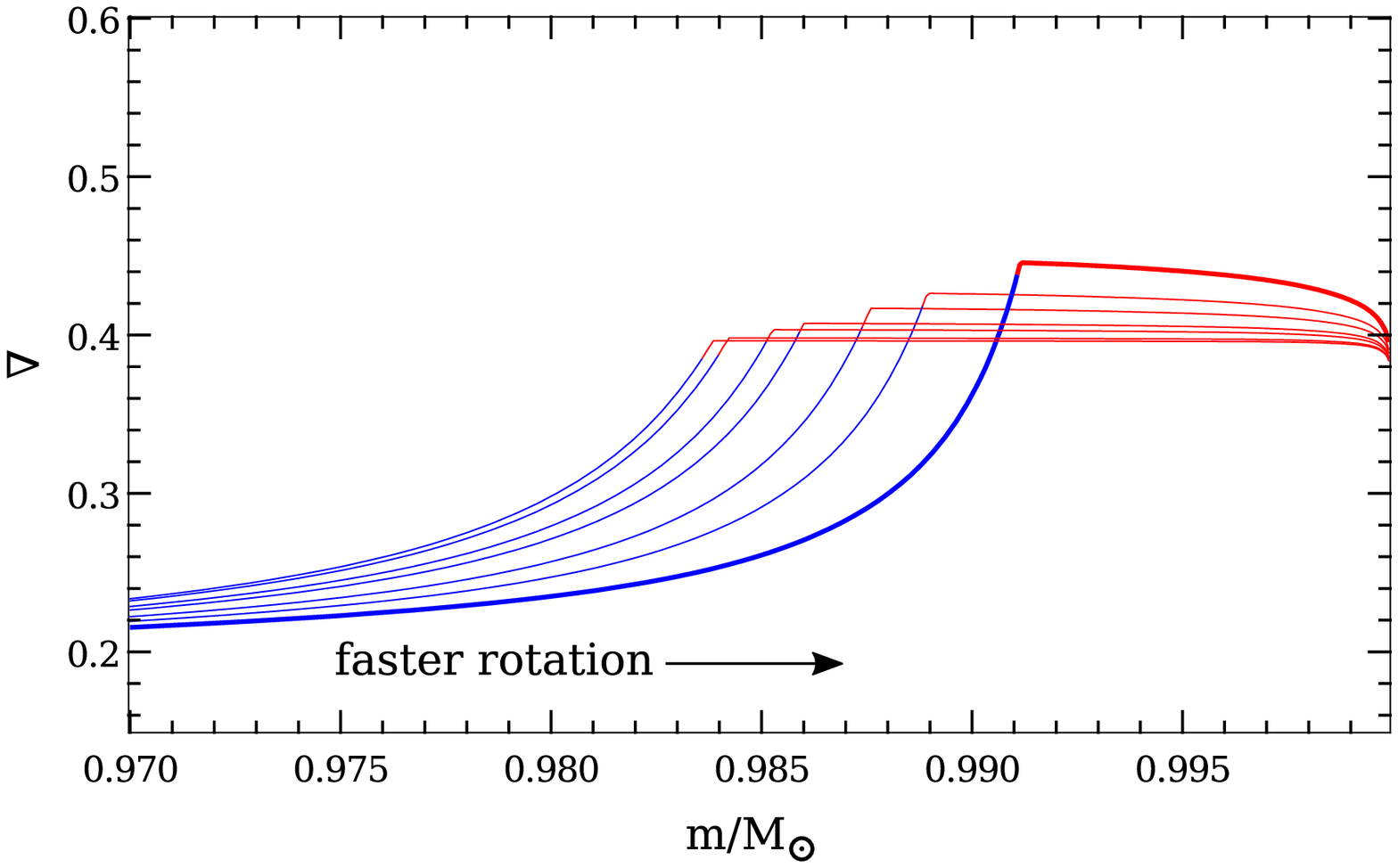}
\newline
\includegraphics[width=\linewidth]{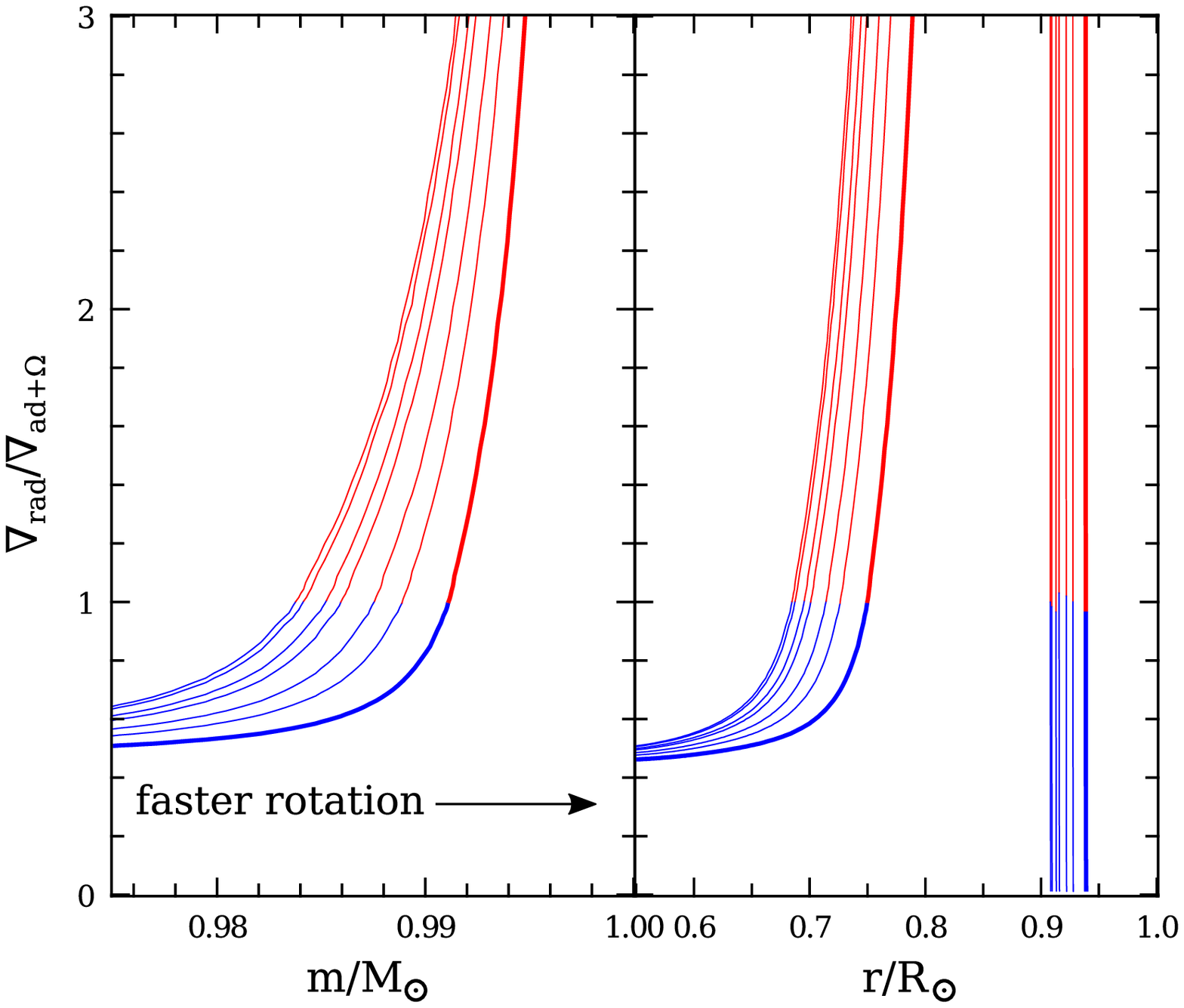}

  \caption{Internal structures from the suite of 1\,M$_\odot$ main-sequence models with different rotation rates ($0 \leq \Omega \leq 0.5\Omega_\text{K}$) shown in Table~\ref{table_structure}.  Upper panel: Temperature and density profiles near the bottom of the convective envelope.  The models towards the left-hand side are the fastest rotators.  Middle panel: Temperature gradient $\nabla = \frac{\text{d}\log{T}}{\text{d}\log{P}}$ for the same models.  The curves that reach the highest $\nabla$ are the fastest rotators and the fastest is denoted by a thick line.  
Lower panel: Ratio of the temperature gradients as a function of enclosed mass and radial coordinate.  The denominator in the ratio of the temperature gradients $\nabla_{\text{ad}+\Omega}$ is the right-hand side of inequality~\eqref{eq_stability_criterion} instead of the true adiabatic temperature gradient $\nabla_\text{ad}$.   The requirement for convective instability in the models with rotation is $\nabla_\text{rad}  / \nabla_{\text{ad}+\Omega} > 1$.  The curves furthest to the right are the fastest rotators and the fastest is denoted by a thick line.
Convectively unstable regions are shown in red and stable regions in blue.}
  \label{figure_rotation_rhoT}
  \label{figure_rotation_ratgrads}
\end{figure}

Figure~\ref{figure_rotation_rhoT} shows the temperature-density profiles as well as the temperature gradient of the models in Table~\ref{table_structure}. Convectively unstable regions are shown in red and stable ones in blue.  The differences between the models are most obvious near the bottom of the convection zone where radiation dominates the heat transport.  This contrasts with the ordinary adjustments to convective efficiency via changes to the mixing length parameter, which have a negligible direct effect on the deeper layers.  It is clear that as the rotation rate increases, the radiative zone extends to cooler temperatures.  The steepening of the temperature gradient near the bottom of the convection zone is also apparent.  Despite the large variation in $\nabla$, the maximum MLT convective velocity in each of the model is nearly identical.  We do not expect these effects to be important for helioseismology: At the solar rotation rate, the structural changes to, for example, the depth of the convection zone are small compared with those from uncertainties such as opacity, equation of state, and the treatment of mixing near the convective boundary.

The fastest modelled rotation rate in Table~\ref{table_structure} and Fig.~\ref{figure_rotation_rhoT}, $P \approx 0.2\,\text{day}$, corresponds approximately to the lower end of the period distribution for solar mass stars among similarly aged clusters: Pleiades \citep{2010MNRAS.408..475H}, M50 \citep{2009MNRAS.392.1456I}, and M35 \citep{2009ApJ...695..679M}.  It is evident from the global properties in Table~\ref{table_structure} and the structures in Fig.~\ref{figure_rotation_rhoT} that the effects are small for periods greater than about half a day.   By the age of the Pleiades this is the majority of stars: Only 20 of the 148 stars with Li and rotation determinations in the sample from \citet{2018A&A...613A..63B} have periods of less than 0.5\,day and seven have periods of less than 0.25\,day.  The effects are more critical, however, at early stages, during the PMS, when the angular momentum is higher and the lithium depletion in standard models is much more rapid.  

\setlength{\tabcolsep}{3pt}
\begin{table*}

\begin{center}
  \caption{Properties of 125 Myr old $1\,\text{M}_\odot$ models with different rotation rates.  $T_\text{bce}$ and $M_\text{bce}$ are the temperature and the mass coordinate at the convective-radiative boundary, respectively.  $\Omega_\text{K} = \sqrt{\frac{G M }{R^3}}$ is the Keplerian break-up rate and $g_\text{eff}/g = 1 - \frac{\Omega^2 r^3}{m\text{G}}$ is the equatorial effective gravity expressed as a fraction of the actual gravity at the base of the convection zone.}
  \label{table_structure}
 \footnotesize
\begin{tabular}{lccccccc}
\toprule
$\Omega$ (rad\,s$^{-1})$ & Period (day) & $\Omega/\Omega_\text{K}$ & $R/\text{R}_\odot$  & $T_\text{eff}$ (K) & $T_\text{bce}$ (K) & $M_\text{bce}/\text{M}_\odot$   & $g_\text{eff}/g$\\
\midrule
0.0                   & -     & 0.000 & 0.9086 &  5792 &  2.0019 &  0.9838 &    1.00 \\
0.721$\times 10^{-4}$ & 1.009 & 0.100 & 0.9098 &  5787 &  1.9854 &  0.9841 &    0.995 \\
1.436$\times 10^{-4}$ & 0.506 & 0.200 & 0.9134 &  5775 &  1.9355 &  0.9852 &    0.982 \\
1.789$\times 10^{-4}$ & 0.406 & 0.250 & 0.9160 &  5766 &  1.9010 &  0.9860 &    0.971 \\
2.388$\times 10^{-4}$ & 0.304 & 0.337 & 0.9217 &  5747 &  1.8209 &  0.9875 &    0.946 \\
2.820$\times 10^{-4}$ & 0.258 & 0.401 & 0.9268 &  5730 &  1.7474 &  0.9889 &    0.921 \\
3.479$\times 10^{-4}$ & 0.209 & 0.502 & 0.9361 &  5699 &  1.6145 &  0.9911 &    0.869 \\
\bottomrule
\end{tabular}
\normalsize
\end{center}
\normalsize
\end{table*}

\section{Lithium evolution effects}
\label{sec:lithium_evolution_effects}

\begin{figure}

\includegraphics[width=\linewidth]{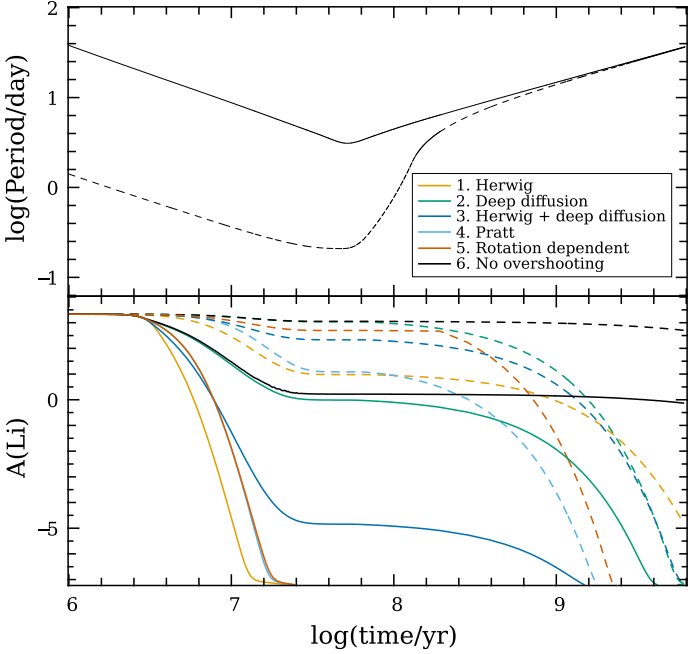}
  \caption{Upper panel: Rotation evolution for representative 0.75\,M$_\odot$ fast (dashed) and slow (solid line) rotators.  Lower panel: Surface lithium evolution for the same fast (dashed) and slow (solid line) rotators with six different treatments of mixing near convective boundaries and the modified stability criterion.}
  \label{figure_lithium_and_rotation_evolution}
\end{figure}

\begin{figure}
\includegraphics[width=\linewidth]{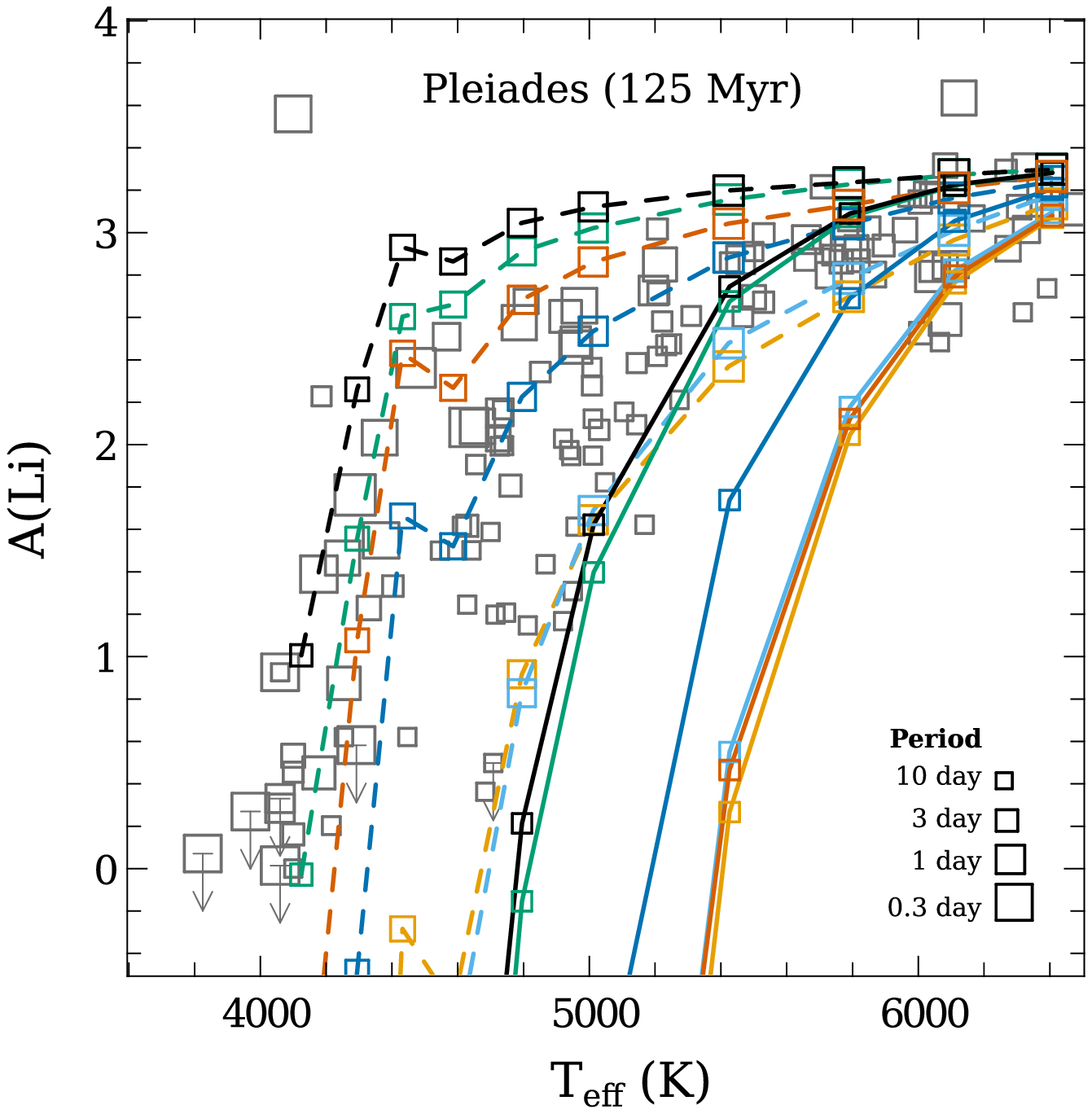}
  \caption{Comparison between observations and theoretical predictions of lithium and rotation in the Pleiades.  Observational data are denoted by grey markers and are taken from \citet{2016A&A...590A..78B}, where the rotational periods are from \citet{2016AJ....152..113R} and lithium abundances are from \citet{2016A&A...596A.113B}.  Marker size is proportional to the logarithm of the rotation frequency.  Theoretical models are fast (dashed lines) and slow (solid lines) rotators with the modified stability criterion and six different treatments of mixing near convective boundaries, which are denoted by the same line colours as in Fig.~\ref{figure_lithium_and_rotation_evolution}.}
  \label{figure_Pleiades_rotation}
\end{figure}

\begin{figure*}
\centering
  \begin{tabular}{@{}ccc@{}}
\includegraphics[width=.300\textwidth]{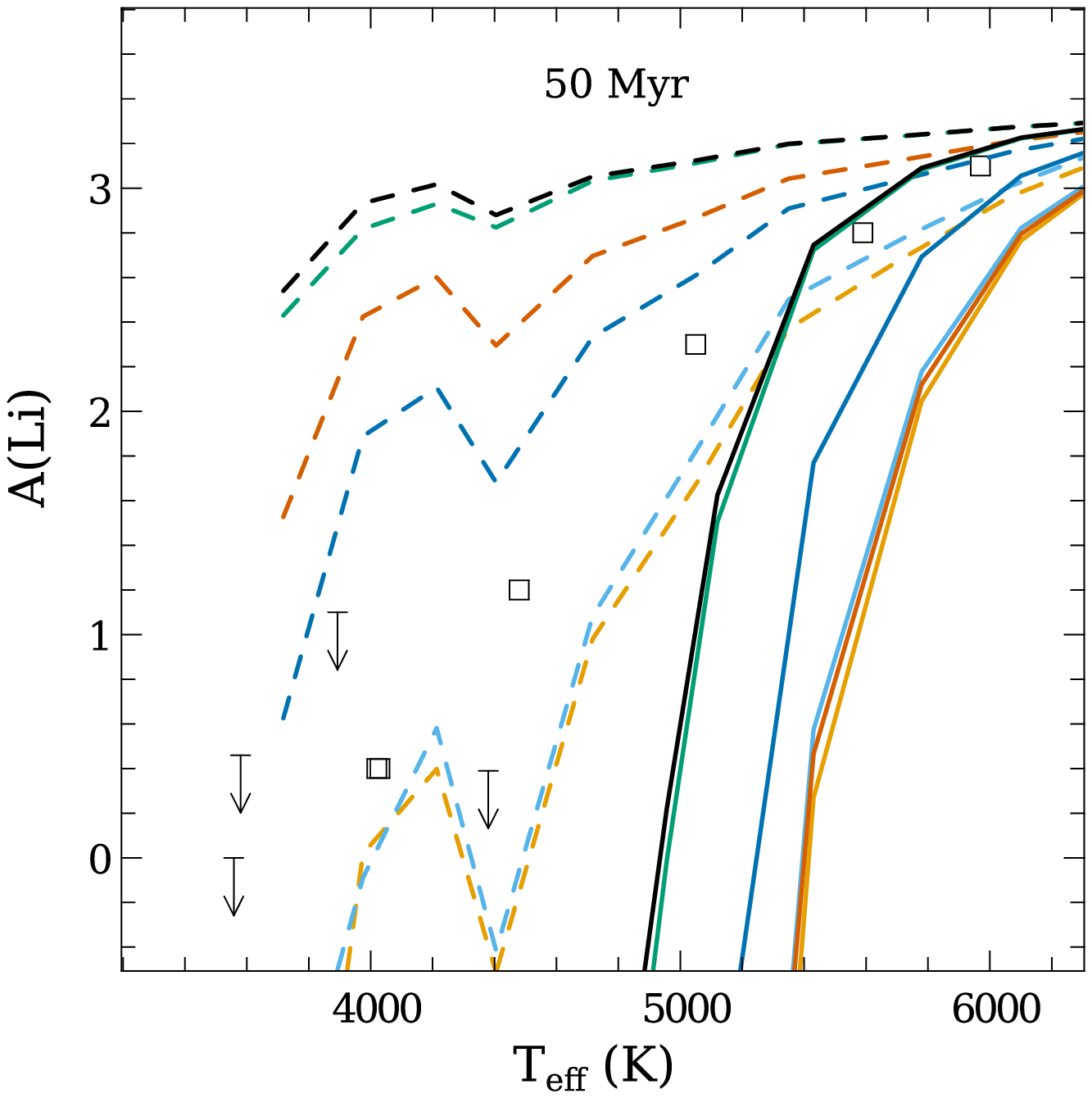}
\hspace{0.3cm} \includegraphics[width=.312\textwidth]{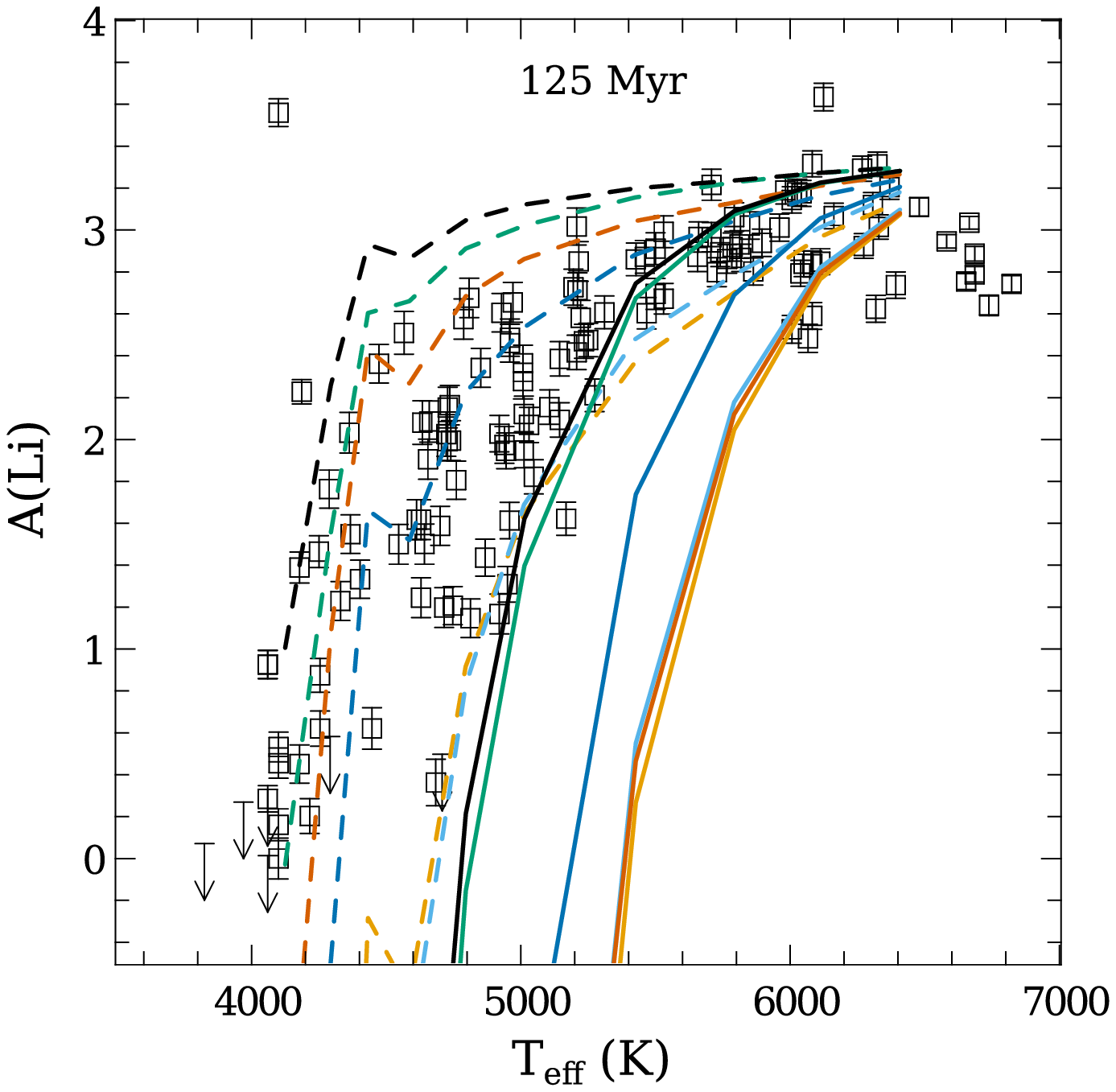}
\includegraphics[width=.300\textwidth]{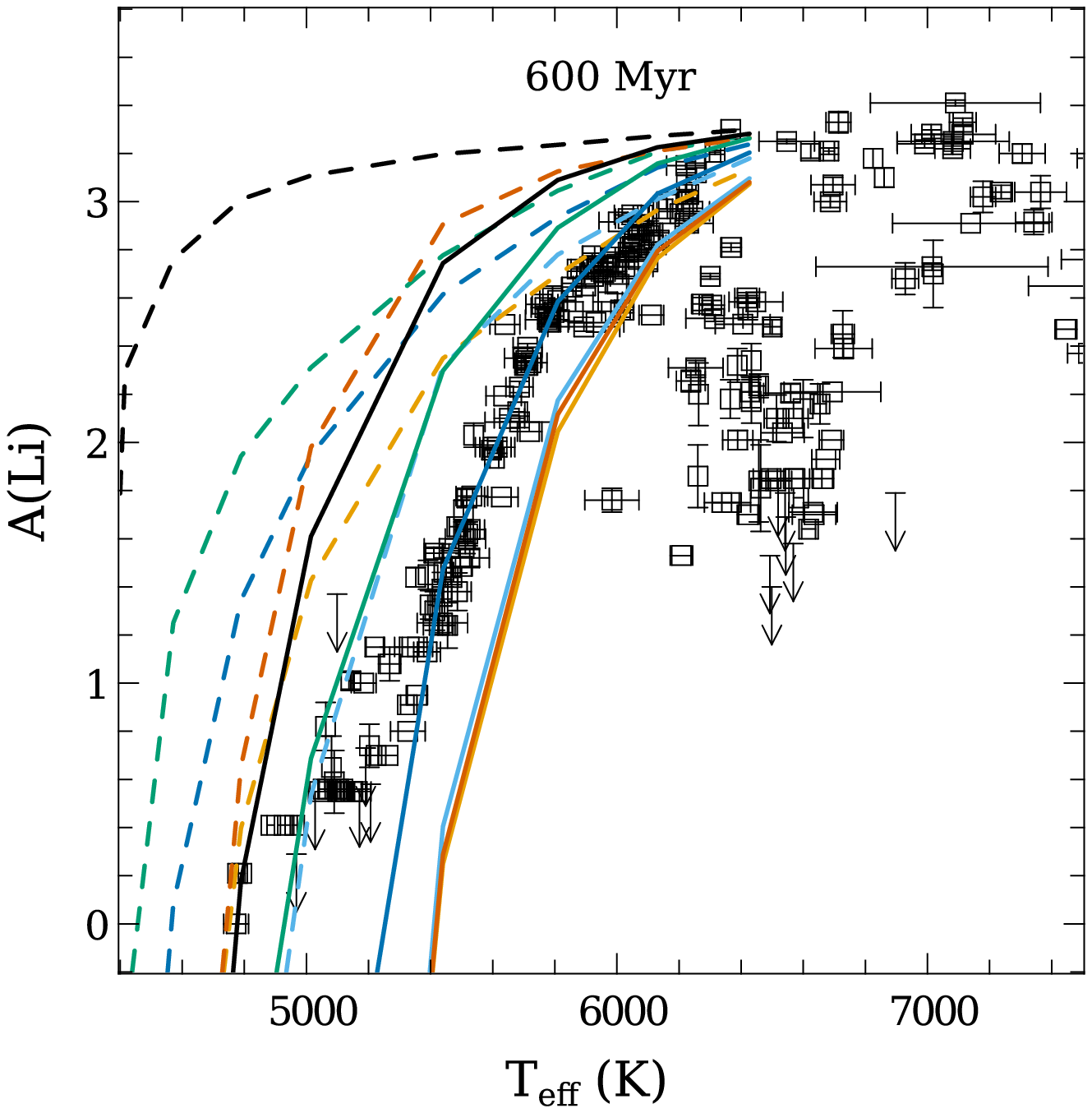}\\
\includegraphics[width=.302\textwidth]{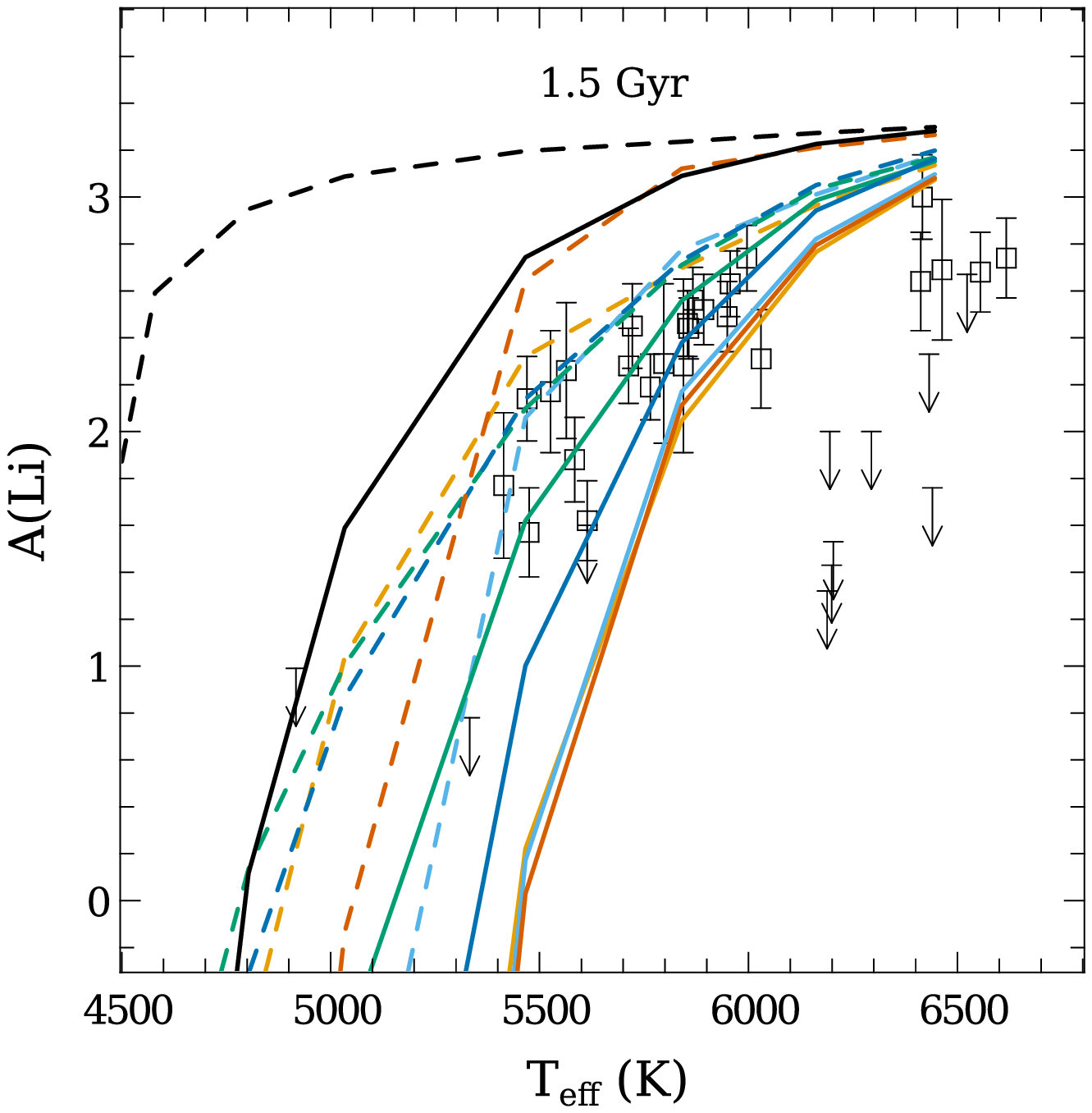}
 \vspace{0.3cm}
\includegraphics[width=.302\textwidth]{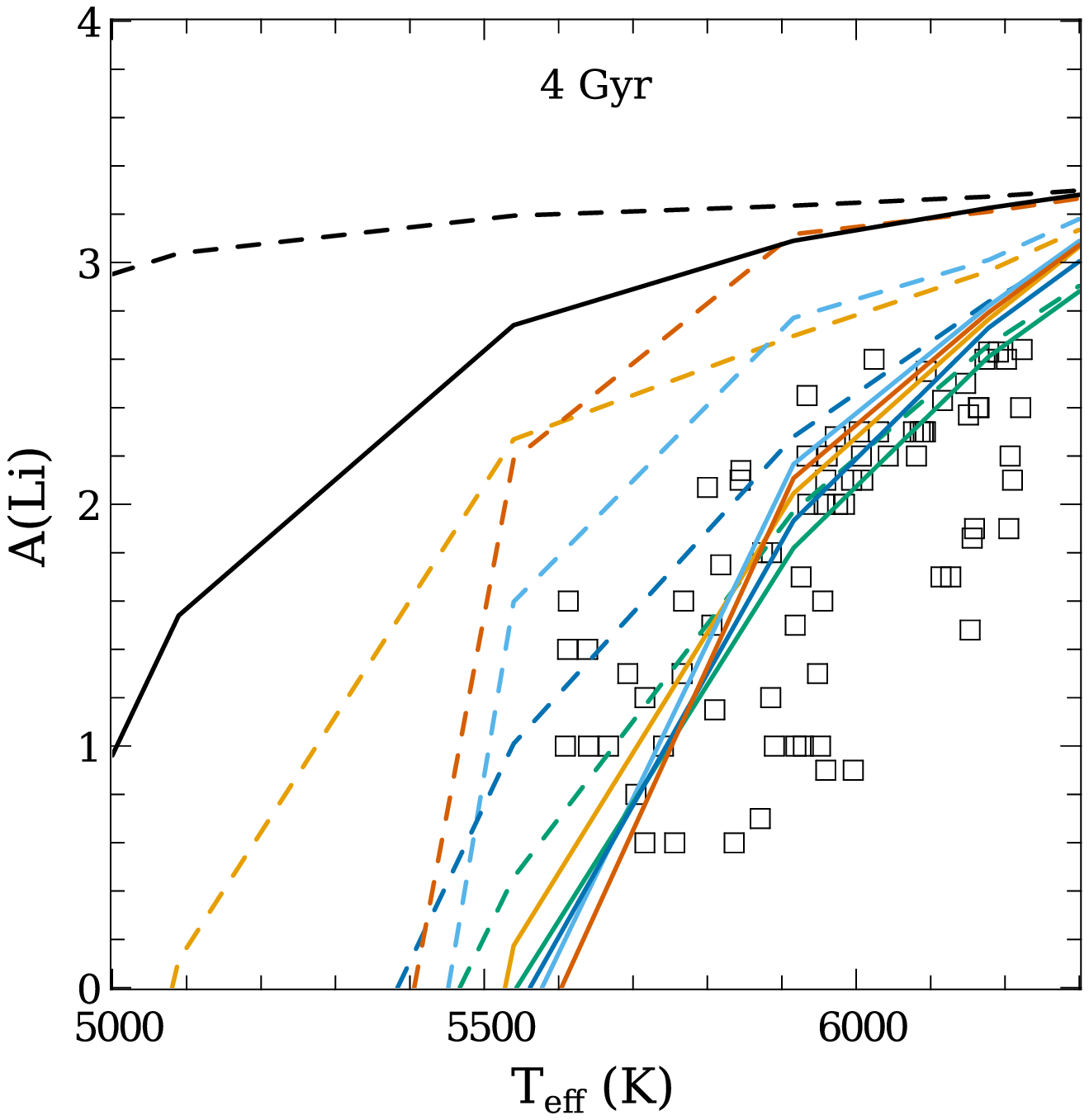}
\includegraphics[width=.293\textwidth]{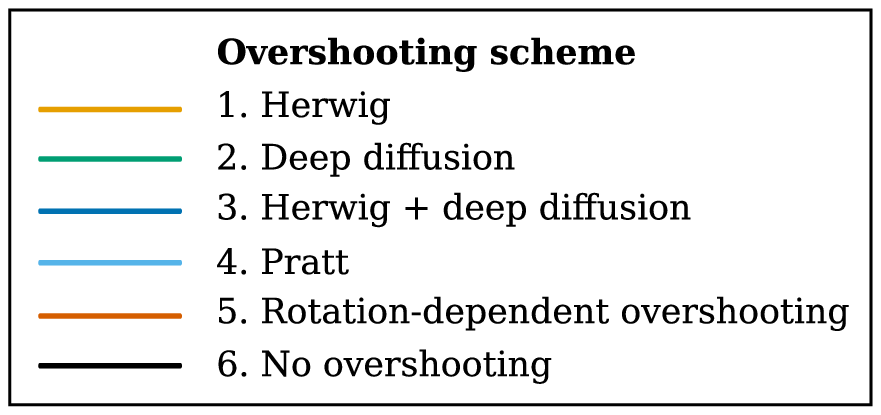}
  \end{tabular}
  \caption{Comparison between the observed lithium abundance in dwarf stars in clusters of various ages and predictions from models with fast (dashed) and slow (solid line) rotation with six different treatments of mixing near convective boundaries and the modified stability criterion.  Data for IC 2391 and IC 2602 (50\,Myr) are from \citet{2001A&A...372..862R}; Pleiades data (125\,Myr) are from \citet{2001A&A...372..862R}; data for the Hyades (600\,Myr) are from \citet{2012A&A...541A.150P} and \citet{2017AJ....153..128C} and for Praesepe (600\,Myr) are from \citet{2017AJ....153..128C}; data for NGC 752 (1.5\,Gyr) are from \citet{2016A&A...590A..94C}; and M67 (4\,Gyr) are from \citet{2012A&A...541A.150P}.  Data shown from \citet{2017AJ....153..128C} is limited to stars for which a $T_\text{eff}$ uncertainty was given. }
  \label{figure_lithium_evolution_clusters}
\end{figure*}

Figure~\ref{figure_lithium_and_rotation_evolution} shows the evolution of the rotation and surface lithium abundance for a set of 0.75\,M$_\odot$ slow and fast rotating models with the modified stability criterion and six different treatments of mixing near convective boundaries outlined in Sect.~\ref{sec_mixing}.  The fast rotation is enough to very strongly suppress PMS lithium depletion in these models.  Models with different mixing schemes are broadly similarly affected by the rotation.  Much of the disparity can be attributed to whether the scheme (and the specific tuning of parameters) favours faster early lithium depletion or slow depletion over the duration of the main sequence.  Those with deep diffusion and to a lesser extent those with Pratt overshooting, which have more main sequence lithium depletion, are the least affected.  The magnitude of the effect from rotation is always significant, however, even for the model without any overshooting rotation causes $\Delta A(\text{Li}) \approx 3$.  The slowly rotating sequences are slow enough that the model with rotation-dependent overshooting is already lithium depleted before the rotation rate exceeds the chosen threshold value ($P = 5\,$day) and the overshooting distance is severely curtailed.

The surface Li evolution is most sensitive to the rotation rate at younger ages when rotation and Li depletion is fastest.  The fast-rotating 0.75\,M$_\odot$ sequence in Fig.~\ref{figure_lithium_and_rotation_evolution} rotates perhaps a little faster than the fastest rotators in Cepheus~OB3b \citep{2010MNRAS.403..545L} but is consistent with those in NGC~2363 \citep{2008MNRAS.384..675I}, both of which are $4 - 5$\,Myr old.  The model is also consistent with the rapid rotators in the roughly 13\,Myr old h~Persei cluster \citep{2013A&A...560A..13M}.  By the age of the Pleiades, the fast-rotating models do not rotate as rapidly as their observed counterparts in clusters: Pleiades \citep{2010MNRAS.408..475H}, M50 \citep{2009MNRAS.392.1456I}, NGC~2516 \citep{2007MNRAS.377..741I}, and M35 \citep{2009ApJ...695..679M}.  Although by this age most stars of this mass have periods of several days and therefore the structure would be unaffected by the rotation.

Lithium predictions from the grid of models at the age of the Pleiades are shown in Fig.~\ref{figure_Pleiades_rotation} along with the observed and theoretical rotation rates.  Irrespective of the mixing scheme, the spread in $A$(Li) due to rotation is at least as large as in the data.  It is not surprising that the slowly rotating models with conventional (or similar) overshooting schemes are too lithium poor at young ages.  It is a well-known issue that ordinary stellar models experience most of the lithium depletion too early if they match the present Sun (and we have calibrated the overshooting to do just that).  Fast rotation for the Herwig and Pratt models is able to suppress lithium depletion enough for them to match the lithium abundances of the observed slow rotators but it should be noted that even a one third increase to the initial rotation rate is enough to shift the curves to higher $A$(Li) so that they fit the upper range of the observations.  At such young ages ($\lesssim 125$\,Myr), more rapid rotation makes $A$(Li) less dependent on the overshooting prescription because very fast rotation will always severely suppress Li depletion.

The marker sizes denote the rotation rates in Fig.~\ref{figure_Pleiades_rotation}, and from this it appears that, if anything, the fast-rotating stars among this group are faster rotators than the models.  Faster rotation would increase $A$(Li), but most of the Li depletion occurs before about 25\,Myr, whereas most of the angular momentum is lost afterwards (Fig.~\ref{figure_lithium_and_rotation_evolution}) and the early rotation rates are consistent with the observations.  This suggests that any rotation rate discrepancy could be a consequence of the rotation model, rather than an indication that the early rotation is too slow.  The predicted rotation rates at older ages will be especially sensitive to the braking law adopted.  We used a single value of $K$ and $\Omega_\text{sat}$ in Eq.~\eqref{eq_angmom} across the stellar mass range studied.  When the \citet{1988ApJ...333..236K} braking law is calibrated using the Sun, the spin-down time is much faster in the saturated regime for stellar masses lower than that of the Sun compared with other braking laws \citep{2010ApJ...722..222B,2012ApJ...746...43R,2013ApJ...776...67V,2015ApJ...799L..23M}.  By increasing the spin-down time in the saturated regime, more recent braking laws would help  reconcile the rotation discrepancy between the lower-mass fast-rotating models and most rapidly rotating stars which is already evident by the age of the Pleiades.  We also did not model a disk-locking phase.  In our solid-body rotation models we do not expect this to have a significant impact on our results because (i) we calibrated the initial rotation to match observed stars, (ii) the phase would be relatively short for the fast rotators, and (iii) it would only affect the rotation rate before lithium depletion is strongest.  In models that allow differential rotation, however, the disk-locking treatment could be important.

Figure~\ref{figure_lithium_evolution_clusters} shows a comparison of Li from the evolution models and various open clusters spanning ages between 50\,Myr and 4\,Gyr.  The fast and slow rotating sequences have $ 0.18 < \Omega/\Omega_\text{K} < 0.48$ and $0.015 < \Omega/\Omega_\text{K} < 0.045 $ at 50\,Myr, respectively, where $\Omega_\text{K}$ is the Keplerian break-up rate.  Among the former fast-rotating models, those with $M \ge 0.75\,\text{M}_\odot$ have $ 0.34 < \Omega/\Omega_\text{K} < 0.48$ while the others have $ \Omega/\Omega_\text{K} < 0.37$.  The sensitivity of the structure to changes in the rotation rate when $\Omega$ is high, which we demonstrated in Fig.~\ref{figure_rotation_rhoT}, means that the abundances for the fast-rotating models shown by the dashed lines should be considered indicative of the magnitude of the effect of rotation, that is small changes to the rotation will move the curves up or down, particularly for the cooler (lower-mass) models where lithium depletion is fastest.  The non-monotonicity of the $A(Li)$ curves for the fast-rotating 50\,Myr models in Fig.~\ref{figure_lithium_evolution_clusters} is due to this sensitivity: Increasing by one third the initial rotation rate of the 0.75\,M$_\odot$ fast-rotating model with Herwig overshooting increases the surface lithium by $\Delta A(\text{Li}) = 1.7$ at 50\,Myr.  That being said, it is quite clear that the observational $A$(Li) data generally sits between the averages of the theoretical curves from the respective rapidly and slowly rotating sequences.   An exception to this is the 4\,Gyr old models where the slow rotators (except for those without overshooting) provide a good match to the data (which has considerable scatter).  This was achieved by construction, however, because the mixing parameters were calibrated to produce the solar lithium value with solar rotation at the solar age and $T_\text{eff}$.

The theoretical curves of $A$(Li)-$T_\text{eff}$ at the lowest $T_\text{eff}$ appear steeper than the relationship implied by the Pleiades (125\,Myr) and the Hyades+Praesepe (600\,Myr) datasets.  A number of the Li determinations are upper limits, which tends to exacerbate the appearance of a disagreement.  The slope of the theoretical curves is also sensitive to the metallicity of the models \citep[see e.g.][]{1997ARA&A..35..557P}.  Higher metallicity sequences tend to experience more lithium depletion, especially at cooler temperatures, producing a steeper $A(\text{Li})$-$T_\text{eff}$ isochrone.  Our models all have solar metallicity so they are consistent with the Pleiades \citep{2009AJ....138.1292S} but the Hyades and Praesepe are both more metal rich \citep{2017AJ....153..128C}, so metallicity is an unlikely culprit.  In addition to the uniform metallicity, the MLT mixing length parameter and the overshooting parameters were each calibrated to the match the (hotter) Sun, so perfect agreement for cooler stars is not expected.

In general, a spread in $A$(Li) between fast and slowly rotating models develops early and then after both models are effectively slow rotators (see e.g. Fig.~\ref{figure_lithium_and_rotation_evolution}) the Li spread is preserved for the rest of the evolution.  This may not be the case, however, with a more sophisticated treatment of rotation (see e.g. the models from \citealt{2021A&A...646A..48D} where the impact of various treatments for rotational mixing as well as transport processes for chemicals and angular momentum are explored).  We deliberately chose extremes in rotation rates to highlight the contrast so it is not surprising that the $A$(Li) spread from our sets at 600\,Myr (and perhaps 50\,Myr and 1.5\,Gyr) is greater than in the data.  This is especially evident for 600\,Myr old stars between about 5400\,K and 5900\,K. Unfortunately, cooler stars ($T_\text{eff} < 6000\,\text{K}$) of that age, in the Hyades for example, are all slow rotators \citep{2013PASJ...65...53T}, so it would be difficult to infer anything about their rotational history and any effect on lithium.

The modified stability criterion and temperature gradient have the least effect on models that experience the most mixing at older ages when rotation is slower, particularly the models with deep diffusion and to a lesser extent the Pratt scheme, which, when calibrated to the Sun, gives more mixing during the main sequence than the Herwig scheme.  The effect is pronounced enough, however, for the rapidly rotating models to usually be more lithium rich as the observed stars (with the exception of the Herwig and Pratt schemes at young ages; $\leq 125$\,Myr).  The young (less than $600$\,Myr) fast-rotating models with rotationally dependent overshooting experience the most suppression of lithium depletion: This is because the overshooting scheme already provides a mechanism to reduce the depth of chemical mixing when rotation is fast, and this works in concert with the structural changes we are testing.  The slowly rotating models with rotation-dependent overshooting have more lithium depletion than those from \citet{2017ApJ...845L...6B}.  This is because we chose more slowly rotating representative models than \citet{2017ApJ...845L...6B}, which consequently spend less time rotating faster than the threshold rotation rate above which overshooting is suppressed: The earlier results from \citet{2017ApJ...845L...6B} could be recovered by choosing a faster initial rotation rate or a lower $\Omega_t$ parameter.

The overall best fitting set of models is probably the one with a combination of an exponential decay in the diffusion coefficient and a minimum diffusion coefficient in the radiative zone (`Herwig + deep diffusion' in Fig.~\ref{figure_lithium_evolution_clusters}), which unsurprisingly predict $A$(Li) somewhere between models with each of those two treatments separately.  These models were computed in order to mimic, at least superficially, those in the literature that are most consistent with the empirical lithium depletion rate over the main sequence.  We note that the Herwig + deep diffusion scheme was only calibrated for the Sun using a single parameter so it could easily be further refined, but we should emphasize that we are not suggesting that this is the correct physical description, nor do we propose a specific physical mechanism that would produce it.  We further add that the choice of combining the deep diffusion scheme with the Herwig scheme was arbitrary: The same effect could be achieved with for example a Pratt + deep diffusion scheme because the crucial aspect is addition of the slow mixing at great depth from the imposition of a minimum diffusion coefficient.

\section{Summary and conclusions}

Inspired by 3D hydrodynamical simulations of rotating convection, we tested how a relatively straightforward change to the stability criterion for convection affects theoretical predictions from 1D models for the surface lithium abundance in low-mass stars.  While integrating this into a stellar evolution code with a more sophisticated treatment of rotation is reserved for future work, we briefly summarize the effects we found.  

We computed a grid of evolution models between 0.55\,M$_\odot$ and 1.2\,M$_\odot$ with representative slow and fast rotation rates, which during the epoch of fast lithium depletion span the range of rotation periods observed in young stellar clusters.  By suppressing lithium depletion in rapid rotators mostly at young ages, these new models reproduce the observed lithium-rotation correlation and help to resolve the problem that conventional stellar models with convective overshooting predict too much Li depletion during the PMS and too little during the main sequence.  The changes to the evolution only become important when rotation is fast ($P \lesssim 0.5$\,day).  The models become very slightly inflated and cooler, but more significantly, the convection zone becomes shallower  and the temperature at its base is reduced as a result.  This moderates the rate of lithium destruction to such an extent that the potential spread in $A$(Li) due to rotation for models of a given effective temperature is at least as large as the observational spread for all of the clusters examined.  The presence of these effects, and to some extent their magnitude, is independent of the choice of prescription for chemical mixing beneath the convection zone. 

The changes to the structure--radial expansion and cooling at the base of the convective envelope--are similar to those that result from other factors, such as magnetic inhibition of convection and spot coverage, which have been proposed to explain the empirical lithium-rotation correlation.  We expect that the modification adopted in this study would generally work in concert with those processes and make it easier to reproduce the observed trend.  In the mass range we examined, the changes to the stellar structure are only significant at very young ages, when rotation is fast, so the structure is essentially unchanged for most of the stellar lifetime and they therefore would probably not be detectable from colour-magnitude diagrams.  

The modification of the stability criterion we implemented can only reduce the rate of lithium depletion and therefore it does not eliminate the need for a process that mixes material beneath the convective envelope, such as convective overshooting.  It is possible that the changes to the conditions at the base of the convection zone would affect different overshooting models in substantially different ways; we found this to some extent for even the simple models we tested, but the general trends were similar.  \citet{2017ApJ...845L...6B} were able to reproduce the observed lithium-rotation trend by severely restricting the extent of overshooting when the rotation rate was above a defined threshold level.  Our models, including those without mixing schemes that explicitly depend on rotation exhibited the same trend.  The sharp dependence on rotation rate occurs because the changes to the temperature gradient and temperature at the base of the convective envelope are both proportional to $\Omega^2$ while the $ ^7\text{Li}(\text{p},\alpha)^4\text{He}$ reaction rate is roughly proportional to $T^{20}$.  The modified stability criterion and temperature gradient may thus provide a mechanism for the depth of mixing near the bottom of the convective boundary to be strongly curtailed when rotation is fast.

This work is the first study of its kind and a number of improvements are possible.  We implemented a simplified form of the stability criterion, whereas the stability is actually latitude-dependent.  Its form in a 1D stellar evolution code could be refined by further detailed 2D or 3D calculations.  We did not construct the suite of models to specifically match the observations of each cluster.  All of the models have the same initial composition, and MLT mixing length parameter and the mixing schemes (where applicable) were calibrated for the Sun.  We adopted a very simple treatment of rotation where it is solid-body and there is no reduction to the effective gravity.  The initial rotation rates for the sequence of fast-rotating models were not adjusted to better match the upper bounds of $A$(Li), although this would be possible because of the strong dependence of $A$(Li) on the rotation rate.  The models neglected any effects of magnetism and spots on convection.  Finally, we also assumed cluster ages from the literature rather than seeking a best fit.  We consciously tested a wide variety of treatments for mixing near the convective boundary and the differences between their outcomes were significant enough to show that careful attention must be paid to this in follow-up studies.

An obvious next step is to integrate this work into stellar evolution codes with standard treatments for angular momentum transport.  When solid-body rotation is assumed, the change to the temperature gradient is proportional to $\Omega^2$ so the findings will be sensitive to the treatment of rotation, and especially so when the rotation is fast.  When solid-body rotation is not assumed it will additionally depend on the angular momentum gradient (inequality~\ref{modified_stability_N2}).  The evolution will also similarly depend on the angular momentum loss law adopted, particularly if it affects the rotation rate during the PMS when lithium depletion is fastest.  It would be interesting to see how the modified temperature gradient affects the predictions from models with rotational mixing schemes such as the recent ones from \citet{2021A&A...646A..48D}.  The mass range could be expanded to study whether there are any implications for models of F-stars, which exhibit the so-called lithium dip in (seen very clearly around 6500\,K in the third panel of Fig.~\ref{figure_lithium_evolution_clusters}) where the thickness of the convection zone is important \citep{1998A&A...335..959T}.  Many of these types of stars are also fast rotators \citep[e.g.][]{1998A&A...335..183Q,2019AJ....158..163D}.

This work is an example of the capacity for multi-dimensional hydrodynamical simulations to inform the ingredients of stellar evolution models and help reconcile theory and observations. It demonstrates the potential of this approach for improving the predictive power of 1D stellar evolution models.

\section*{Acknowledgements}

This project was supported by the European Research Council through ERC AdG No. 787361-COBOM and the STFC Consolidated Grant ST/R000395/1.  The authors would like to acknowledge the use of the University of Exeter High-Performance Computing (HPC) facility ISCA and of the DiRAC Data Intensive service at Leicester, operated by the University of Leicester IT Services, which forms part of the STFC DiRAC HPC Facility. The equipment was funded by BEIS capital funding via STFC capital grants ST/K000373/1 and ST/R002363/1 and STFC DiRAC Operations grant ST/R001014/1. DiRAC is part of the National e-Infrastructure.

\footnotesize{
 \bibliographystyle{aa}
  \bibliography{refs}
}

\end{document}